\documentclass{ws-ijmpa}
\newcommand{\tr}{\rm tr \,}
\begin{document}

\markboth{Authors' Names}
{Instructions for Typing Manuscripts (Paper's Title)}

%
\catchline{}{}{}{}{}
%

\title{DYNAMICS OF STRONG AND RADIATIVE DECAYS\\
OF D$_s$ MESONS
}
\author{MADELEINE SOYEUR
}

\address{Irfu/SPhN, CEA/Saclay\\ F-91191 Gif-sur-Yvette Cedex, France\\
madeleine.soyeur@cea.fr}

\author{MATTHIAS F. M. LUTZ}

\address{GSI, Planckstrasse 1\\
D-64291 Darmstadt, Germany\\
m.lutz@gsi.de}

\maketitle

\begin{history}
\received{Day Month Year}
\revised{Day Month Year}
\end{history}

\begin{abstract}
The positive parity scalar D$_{s0}^*$(2317) and axial-vector D$_{s1}^*$(2460)
charmed strange mesons are generated by coupled-channel dynamics through the s-wave scattering of Goldstone bosons off the pseudoscalar and vector D(D$_s$)-meson ground states. The attraction lea\-ding to the specific masses of these states reflects the chiral symmetry breaking scale which characterizes the Weinberg-Tomozawa interaction in the chiral Lagrangian. Chiral corrections to order Q$_\chi^2$ are calculated and found to be small. The D$_{s0}^*$(2317) and D$_{s1}^*$(2460) mesons decay either strongly into the isospin-violating $\pi^0$D$_s$ and $\pi^0$D$_s^*$ channels or electromagnetically. We show that the $\pi^0$-$\eta$ and (K$^0$D$^+$-K$^+$D$^0$) mixings act constructively to generate strong widths of the order of 140 keV. The one-loop contribution to the radiative decay amplitudes of scalar and axial-vector states is calculated using the electromagnetic Lagrangian to chiral order Q$_\chi^2$. We show the importance of taking into account processes involving light vector mesons explicitly in the dynamics of electromagnetic decays to obtain a satisfactory description of the available data.

\keywords{Charmed mesons; D$_{s0}^*$(2317); D$_{s1}^*$(2460).}
\end{abstract}

\ccode{PACS numbers: 11.10.St; 12.39.Fe; 13.20.Fc.}

\section{Introduction}	

The positive parity scalar D$_{s0}^*$(2317) and axial-vector D$_{s1}^*$(2460)
charmed strange mesons have peculiar properties. Their masses are significantly lower than expected in quark models \cite{Godfrey}. Their strong decays into the $\pi^0$D$_s$ and $\pi^0$D$_s^*$ channels violate isospin
and their radiative decays appear quite selective.
The present upper limit for the  D$_{s0}^*$(2317) total width is $\Gamma < 3.8$ MeV \cite{Aubert2}. Its radiative width should be extremely small according to the constraint \cite{PDG:2006}
\begin{eqnarray}
\frac{\Gamma \,[D_{s0}^*(2317) \, \rightarrow \,  D_{s}^*(2112)\, \gamma]}
{\Gamma\,[D_{s0}^*(2317)\, \rightarrow \, D_{s}(1968)\, \pi^0]}
<\, 0.059.
\label{Dstar0widthtoDstarfirst}
\end{eqnarray}
\noindent
The D$_{s1}^*$(2460)-meson's total width is less than 3.5 MeV
\cite{Aubert2}. Constraints on its radiative decays are as follows \cite{PDG:2006},
\begin{eqnarray}
\frac{\Gamma \,[D_{s1}^*(2460) \, \rightarrow \,  D_{s}(1968)\, \gamma]}
{\Gamma\,[D_{s1}^*(2460) \, \rightarrow \, D_{s}^*(2112)\, \pi^0]}
=0.31 \pm 0.06,
\label{Dstar1widthtoDsfirst}
\end{eqnarray}
\begin{eqnarray}
\frac{\Gamma \,[D_{s1}^*(2460) \, \rightarrow \,  D_{s}^*(2112)\, \gamma]}
{\Gamma\,[D_{s1}^*(2460) \, \rightarrow \, D_{s}^*(2112)\, \pi^0]}
<\,0.16,
\label{Dstar1widthtoDstarfirst}
\end{eqnarray}
\begin{eqnarray}
\frac{\Gamma \,[D_{s1}^*(2460) \, \rightarrow \,  D_{s0}^*(2317)\, \gamma]}
{\Gamma\,[D_{s1}^*(2460) \, \rightarrow \, D_{s}^*(2112)\, \pi^0]}
<\,0.22.
\label{Dstar1widthtoDstar0first}
\end{eqnarray}
The purpose of the work \cite{Lutz-Soyeur-2007} summarized in this paper is to provide a consistent explanation of these properties of the
D$_{s0}^*$(2317) and D$_{s1}^*$(2460) mesons in the hadrogenesis conjecture \cite{Lutz-Kolomeitsev-2001,Lutz-Kolomeitsev-2002} extending earlier investigations \cite{Lutz-Kolomeitsev-2004,Kolomeitsev-Lutz-2004,Hofmann-Lutz-2004} restricted to the study of the masses of these states.

We recall briefly in Section 2 the main assumptions underlying the hadrogenesis conjecture. We outline in Sections 3 how the strong and electromagnetic decays of the D$_{s0}^*$(2317)and D$_{s1}^*$(2460) mesons are calculated and mention the dynamical origin of the features summarized above. We refer to Ref.~4 for a complete presentation of the formalism and of the numerical results.

\section{Hadrogenesis}

The hadrogenesis conjecture \cite{Lutz-Kolomeitsev-2001,Lutz-Kolomeitsev-2002} applied
to heavy-light meson states \cite{Kolomeitsev-Lutz-2004,Hofmann-Lutz-2004}
has been successful in generating hadron resonances
by coupled-channel dynamics with $J^\pi$ = $0^+,\,1^+$ quantum numbers. For D$_s$-mesons, they are produced by the s-wave scattering of Goldstone bosons off D-meson ground state triplets. The D-meson ground state triplet is composed either of the pseudoscalars $\{D^0(1864), -D^+(1869), D_s^+(1968)\}$
to build the $0^+$ state or of the vectors $\{D^{*0}_\mu(1864), -D^{*+}_\mu(1869), D_{s\mu}^{*+}(1968)\}$
to generate the $1^+$ state. For the
D$_{s0}^*$(2317)$^+$ meson, the calculation involves the
$\eta D_s^+$, $K^0 D^+$ and $K^+ D^0$ channels coupled further to the $\pi^0 D_s^+$ channel through
isospin-mixing parameters. Analogously we consider for the D$_{s1}^*$(2460)$^{+}$ meson the
$\eta D_s^{*+}$, $K^0 D^{*+}$, $K^+ D^{*0}$ and $\pi^0 D_s^{*+}$ channels. Both the D$_{s0}^*$(2317)$^+$
and D$_{s1}^*$(2460)$^{+}$ mesons are SU(3) anti-triplet bound states.

The interaction between the Goldstone bosons and the scalar D-meson triplet is governed by the leading order chiral Lagrangian density \cite{Kolomeitsev-Lutz-2004,Hofmann-Lutz-2004}
\begin{eqnarray}
{\mathcal L} &=& \frac{1}{4} \,{\tr } (\partial_\mu \Phi)\,(\partial^\mu \Phi) -\frac{1}{4}\,\tr
\chi_0\,\Phi^2+\ (\partial_\mu D) \, (\partial^\mu \bar D)- D\,M^2_{0^-}   \,\bar D
\nonumber\\
&+& \frac{1}{8\,f^2}\,\Big\{ (\partial^\mu D)\,
\,[\Phi  , (\partial_\mu \Phi)]_-\,\bar D -D\,
\,[\Phi  , (\partial_\mu \Phi)]_-\,(\partial^\mu \bar D )
 \Big\} \,,
 \label{WT-term-scalar}
\end{eqnarray}
where $\Phi$ and $D $ are the pseudoscalar octet and triplet fields. We use the notation
$\bar D = D^\dagger$. The value of the octet meson decay constant f is taken to be f = 90 MeV.
 The ground-state scalar D-meson mass matrix is denoted by $M_{0^-}$. The mass term of the Goldstone bosons is proportional to the quark-mass matrix $\chi_0$.

 An expression similar to (\ref{WT-term-scalar}) is derived for the interaction between the Goldstone bosons and the vector D-meson triplet,
 \begin{eqnarray}
{\mathcal L}&=&  -(\partial_\mu D^{\mu \alpha})   \,(\partial^\nu \bar D_{\nu \alpha})
+ \frac{1}{2}\,D^{\mu \alpha}\,M^2_{1^-}  \,\bar D_{\mu \alpha}
\nonumber\\
&-& \frac{1}{8\,f^2}\,\Big\{ (\partial^\nu D_{\nu \alpha })\,
\,[\Phi  , (\partial_\mu \Phi)]_-\,\bar D^{\mu \alpha }  - D_{\nu \alpha }\,
\,[\Phi  , (\partial_\nu \Phi)]_-\,(\partial_\mu \bar D^{\mu \alpha } ) \Big\} \,,
 \label{WT-term-tensor-axial}
\end{eqnarray}
where we represent the $1^-$ D-mesons
in terms of antisymmetric tensor fields.
This particular representation has the advantage of leading to gauge-invariant expressions for the radiative processes considered in this work. The results obtained in Refs. 5 and 6 with the vector re\-presentation were reformulated in the tensor representation \cite{Lutz-Soyeur-2007}.
Using these chiral Lagrangians at leading order, strong attraction was found in the chiral SU(3) antitriplet states to which the D$_{s0}^*$(2317) and D$_{s1}^*$(2460) belong. The key parameter underlying this result is the chiral symmetry breaking scale $f$.
To improve on the leading order calculation, chiral correction terms were included in a systematic way to order
Q$_\chi^2$ \cite{Hofmann-Lutz-2004}. They take into account the s- and u-channel exchanges of D-mesons and local counter terms. The corresponding vertices introduce additional coupling constants constrained by data when available, heavy quark symmetry and the large N$_c$ limit of QCD \cite{Lutz-Soyeur-2007}.

\section{Strong and radiative decays of the D$_{s0}^*$(2317) and D$_{s1}^*$(2460)}

The strong decays of the D$_{s0}^*$(2317) and D$_{s1}^*$(2460) into the $\pi^0 D_s$ and  $\pi^0 D_s^*$ respectively violate the isospin symmetry. The origin of these processes is the difference between the up- and down-quark masses which induces two kinds of isospin mixings: the $\pi^0$-$\eta$ mixing characterized by an angle $\epsilon$ (determined to be ~0.01 \cite{Gasser-Leutwyler-1985}) and the ($K^0\,D^+$-$K^+_,D^0$) mixing through I=1 transitions. These effects act constructively~\cite{Lutz-Soyeur-2007}.
The isospin-violating strong widths of the D$_{s0}^*$(2317) and D$_{s1}^*$(2460) mesons are sensitive to both the coupled-channel dynamics and the angle $\epsilon$. We find that the strong widths of the D$_{s0}^*$(2317) and D$_{s1}^*$(2460) are comparable (as suggested by the heavy-quark symmetry of QCD) and of the order of 140 keV \cite{Lutz-Soyeur-2007}. This is smaller than the experimental upper limits by a factor of the order of 20. More accurate determinations of these widths would clearly be very useful in unraveling the dynamics underlying these decays.

To calculate the radiative decays of the D$_{s0}^*$(2317) and D$_{s1}^*$(2460) mesons we define the Lagrangian for electromagnetic interactions to chiral order
Q$_\chi^2$. It involves four contributions. First, we gauge the Lagrangian describing strong interactions. Only the 3-point vertices involved in the chiral correction terms matter. Secondly, we consider terms of chiral order
Q$_\chi^2$ proportional to the electromagnetic tensor $F_{\mu\nu}$ and corresponding to 4-point vertices ($D^*D\pi\gamma$ for example). A third kind of terms of chiral order
Q$_\chi^2$ and proportional to $F_{\mu\nu}$ describes the anomalous processes induced by 3-point vertices ($D^*D\gamma$) and the magnetic moments of the charmed vector mesons. A fourth set of terms is introduced to investigate the role of light vector mesons in the radiative transitions.

\begin{figure}[t]
\centerline{\psfig{file=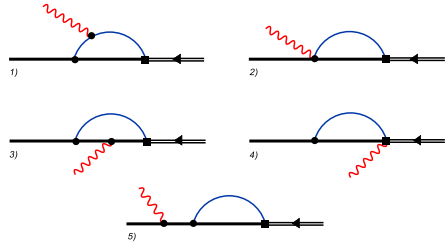,width=5.8cm}}
\vspace*{8pt}
\centering
\caption{Diagrams contributing to the decay amplitude of a scalar
or axial vector molecule.
Solid lines  represent the propagation of
the pseudo-scalar or vector mesons, where the thin lines stand for the light mesons and the thick lines for the heavy
mesons. The wavy line is the photon.} \label{fig}
\end{figure}

We display in Fig. 1 the diagrams contributing to the electromagnetic decay of the D$_{s0}^*$(2317) and D$_{s1}^*$(2460) mesons in the hadrogenesis picture. The evaluation of these graphs is described in Ref. 4. Our numerical results are linked to specific choices of parameters. The explicit introduction of light vector mesons in the radiative decay dynamics in needed to describe the available data summarized in Eqs. 1-4 with natural size parameters. The light vector mesons play a role through gauge-invariant 3- and 4-point vertices proportional to the electromagnetic field strength tensor.

Specific dynamical effects emerge from our work. The full relativistic loop computations induce significant deviations from heavy-quark symmetry as expected from the semi-heavy character of the charm-quark mass. The $\eta\,D_s$ and $\eta\,D_s^*$ channels play an important role in radiative decays. Because of large cancellations between graphs involving $\eta\,D_s$ and $KD$ channels, we are not able to make definite predictions at this point. The investigation of the role of vector mesons indicates that the $K^*D^*$ and $\phi D_s^*$ channels could contribute significantly to the D$_{s0}^*$(2317) and D$_{s1}^*$(2460) radiative decays.

More accurate data, further coupled-channel studies and an improved knowledge of the couplings involved in the calculation (through measurements or lattice studies) are needed to make progress in the dynamical understanding of the structure of the
D$_{s0}^*$(2317) and D$_{s1}^*$(2460) mesons.


\end{document}